\documentclass[pra,twocolumn,superscriptaddress,super,superbib,amsmath]{revtex4-1}
\usepackage{graphicx}
\usepackage[squaren]{SIunits}
\usepackage{color}
\usepackage{ulem}

\def\bra#1{{\langle{#1}|}}
\def\ket#1{{|{#1}\rangle}}

\begin{document}

\title{Deterministic and Robust Entanglement of Nitrogen Vacancy Centers using Low-Q Photonic Crystal Cavities}

\author{Janik Wolters}
\thanks{J. Wolters and J. Kabuss  contributed equally to this work. }
\affiliation{Nano-Optics, Institute of Physics, Humboldt-Universit\"{a}t zu
Berlin, Newtonstr.~15, D-12489  Berlin, Germany}

\author{Julia Kabuss}\email{jkabuss@mailbox.tu-berlin.de}
\affiliation{Institut f\"ur Theoretische Physik, Nichtlineare Optik und Quantenelektronik, Technische Universit\"at Berlin,
Hardenbergstra{\ss}e 36, EW 7-1 10623 Berlin, Germany}

\author{Andreas Knorr}%
\affiliation{Institut f\"ur Theoretische Physik, Nichtlineare Optik und Quantenelektronik, Technische Universit\"at Berlin,
Hardenbergstra{\ss}e 36, EW 7-1 10623 Berlin, Germany}

\author{Oliver Benson}
\affiliation{Nano-Optics, Institute of Physics, Humboldt-Universit\"{a}t zu
Berlin, Newtonstr.~15, D-12489  Berlin, Germany}

\begin{abstract}
We propose an experiment to generate deterministic entanglement between separate nitrogen vacancy (NV) centers mediated by the mode of a photonic crystal cavity.
Using numerical simulations the applicability and robustness of the entanglement operation to parameter regimes achievable with present technology is investigated.
We find that even with moderate cavity Q-factors of $10^{4}$ a concurrence of $c>0.6$ can be achieved within a time of $t_{max}\approx150$~ns, while Q-factors of $10^{5}$ promise $c>0.8$.
Most importantly, the investigated scheme is relative insensitive to spectral diffusion and differences between the optical transitions frequencies of the used NV centers. 
\end{abstract}

\maketitle
Entanglement is one of the most fascinating aspects of quantum mechanics. 
This concept finds application in the field of quantum information processing,  metrology, or secure communication.
Thus, many groups all over the world are striving for realizing entanglement on a large scale.
Although many experiments strikingly demonstrated entanglement of photons~\cite{Aspect1982}, ions~\cite{Blatt2012}, and atoms~\cite{Evellin2010,Gaetan2010,Hofmann2012} these approaches are difficult to scale to a quantum information processing network with many nodes each having several quantum registers~\cite{Kimble2008}.
In contrast, solid state based quantum platforms like quantum dots, superconducting circuits or color centers are in principle scalable.
Among these, the negatively charged nitrogen vacancy (NV) center in diamond is regarded as one of the most promising candidates~\cite{Hanson2008,Jelezko2006, Wrachtrup2006}. 
The NV center provides a triplet ground state with extreme long coherence times, frequently used as spin qubit~\cite{Jelezko2006,Wrachtrup2006,Shi,Hanson2008,Balasubramanian2009,Wolters2013c} and an optical transition at 637~nm suitable to generate narrow band single photons~\cite{Bernien2012}, or to coherently manipulate the NV state~\cite{Batalov2008,Bernien2013}.
Importantly, the NV center also provides a $\Lambda$-type three-level system~\cite{Yale2013a,Santori2006,Togan2010,Maze2011}, which is used for the studied entanglement scheme.
Furthermore, technological progress  of the recent years made it possible to integrate single NV centers into photonic crystal cavities~\cite{Benson2011,Faraon2011a,VanderSar2011,Wolters2010,Wolters2012}.\\
Entanglement between an NV and adjacent nuclear spins~\cite{Fuchs2011}, and two NV centers~\cite{Dolde2013} separated by 25~nm could be achieved using short range spin-spin interactions.
Recently, a probabilistic entanglement scheme~\cite{Barrett2005} could be demonstrated for NV centers being 3~m apart~\cite{Bernien2013}.
The short-range interaction might be suited for quantum registers in a future quantum information processing node, while the probabilistic scheme might be applied to connect different nodes of future quantum information processing networks.
Nevertheless, neither of the demonstrated schemes is suitable for fast operations between several registers in a quantum node on the medium range, i.e. on the order of a wavelength.
In this range an integrated optical platform promises scalability -- at least to the level of several quantum registers --, as well as operations much faster than the coherence time.
\begin{figure}[t]
\center
\includegraphics[width=0.35\textwidth]{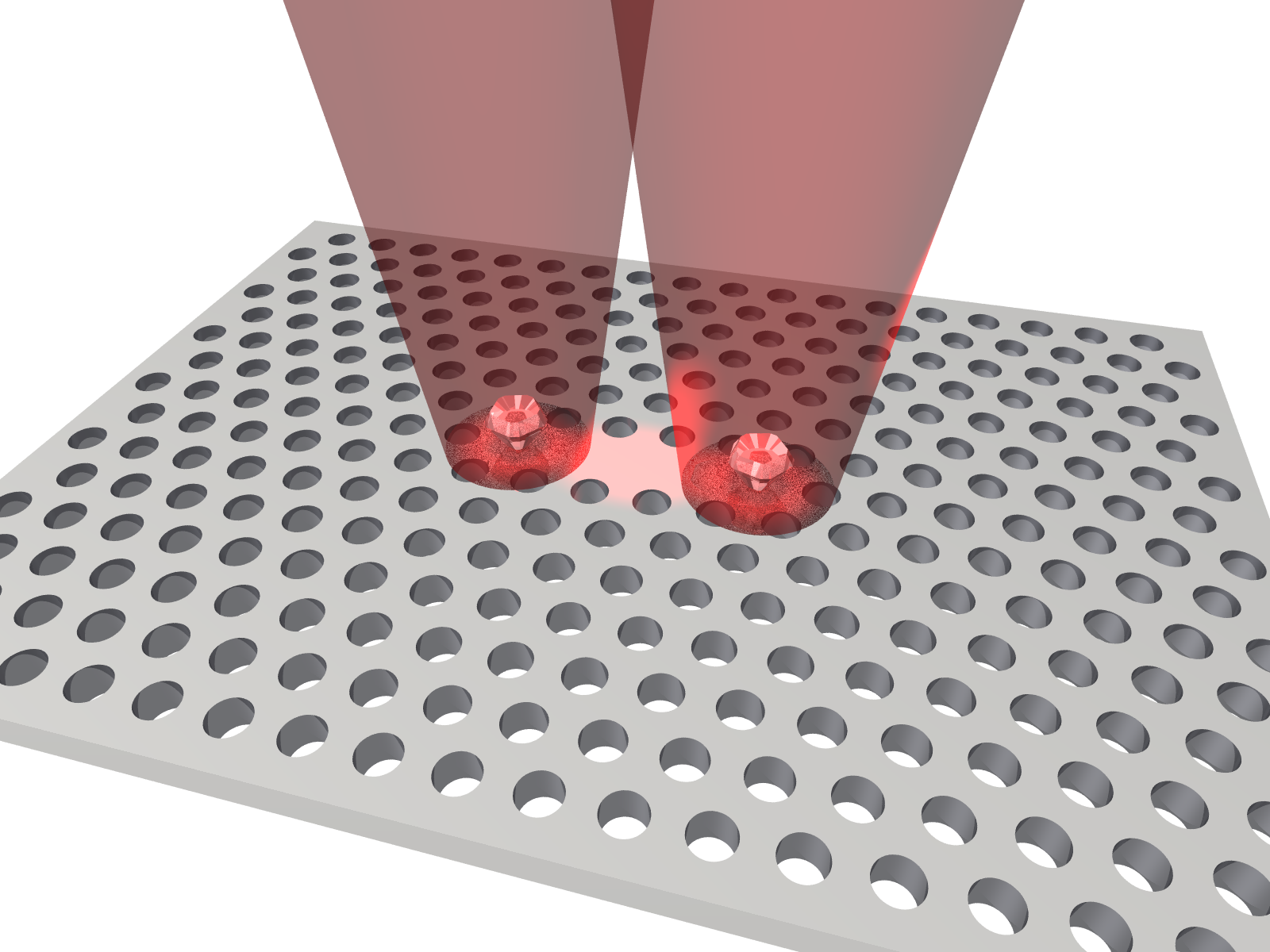}
\caption{(Color online) Artist's view of the considered system of two NV centers in nanodiamonds coupled via a photonic crystal cavity formed by a row of missing holes.}
\label{fig:art_view}
\end{figure}
Recently Yang et al. proposed that an interaction between medium distant NVs can be mediated by high quality cavities with Q-factors exceeding $10^{6}-10^{8}$~\cite{Yang2011,Yang2010a,Yang2010,Liu2013}.
Achieving such high Q-factors in cavities with incorporated NV centers is technologically extremely challenging~\cite{Mccutcheon2008,Riedrich-Moller2011a}.
 In general, a practical protocol to entangle solid-state quantum emitters has to cope with non-identical emitter properties and -- even more importantly -- it has to be robust against fluctuations. 

In this paper we regard an entangling operation that is relative insensitive to differences of the emitters' optical transitions frequencies, that tolerates spectral diffusion, and that requires only experimentally feasible optical cavities with Q-factors of about $10^4- 10^5$.
 Applying numerical simulations we show that entanglement of medium-distant NVs sharing a low-Q mode of a photonic crystal cavity is possible. 
 Although we examplarily regard a specific system here, the scheme is applicable to other types of cavities and quantum systems.
 
In the following we first introduce the model system and compare to related analytical results of Ref. [32]. 
Then, we adopt the model to our realistic scenario with NV centers and a parameter range that has already been achieved in current experiments. 
We numerically solve the equations of motion, showing that the scheme can compete with other entanglement methods.

The key elements of the entanglement scheme are two $\Lambda$-type systems (e.g. NV centers) with long lived spin ground states $\ket 0$, $\ket 1$, in which a qubit can be encoded, and an excited state $\ket E$.
These are placed in two anti-nodes of the mode of a low-Q photonic crystal cavity with small mode volume (Fig.~\ref{fig:art_view}). 
This configuration allows for independent optical initialization and read out of both systems.
Furthermore, coherent all-optical one-qubit operations, e.g. in the Raman scheme~\cite{Moler1992} are possible: 
Two laser fields, one with frequency $\omega$ coupling to the transition $\ket 0 \leftrightarrow \ket E$ with strength $\Omega$, the other with frequency $\omega'$ coupling to $\ket 1 \leftrightarrow \ket E$ with strength $\Omega'$ are applied to an individual system. If  the frequency difference $\delta\omega=\omega-\omega'$ corresponds to the energy spacing $\omega_{01}$ between $\ket 0$ and $\ket 1$ and the lasers are detuned by $\Delta$ from the respective transition to the excited state, the system undergoes a spin rotation with the frequency $\Omega_{Raman}={\Omega\cdot\Omega'}/{(2\Delta)}$.

\begin{figure}[hbt]
\center
\includegraphics[width=0.35\textwidth]{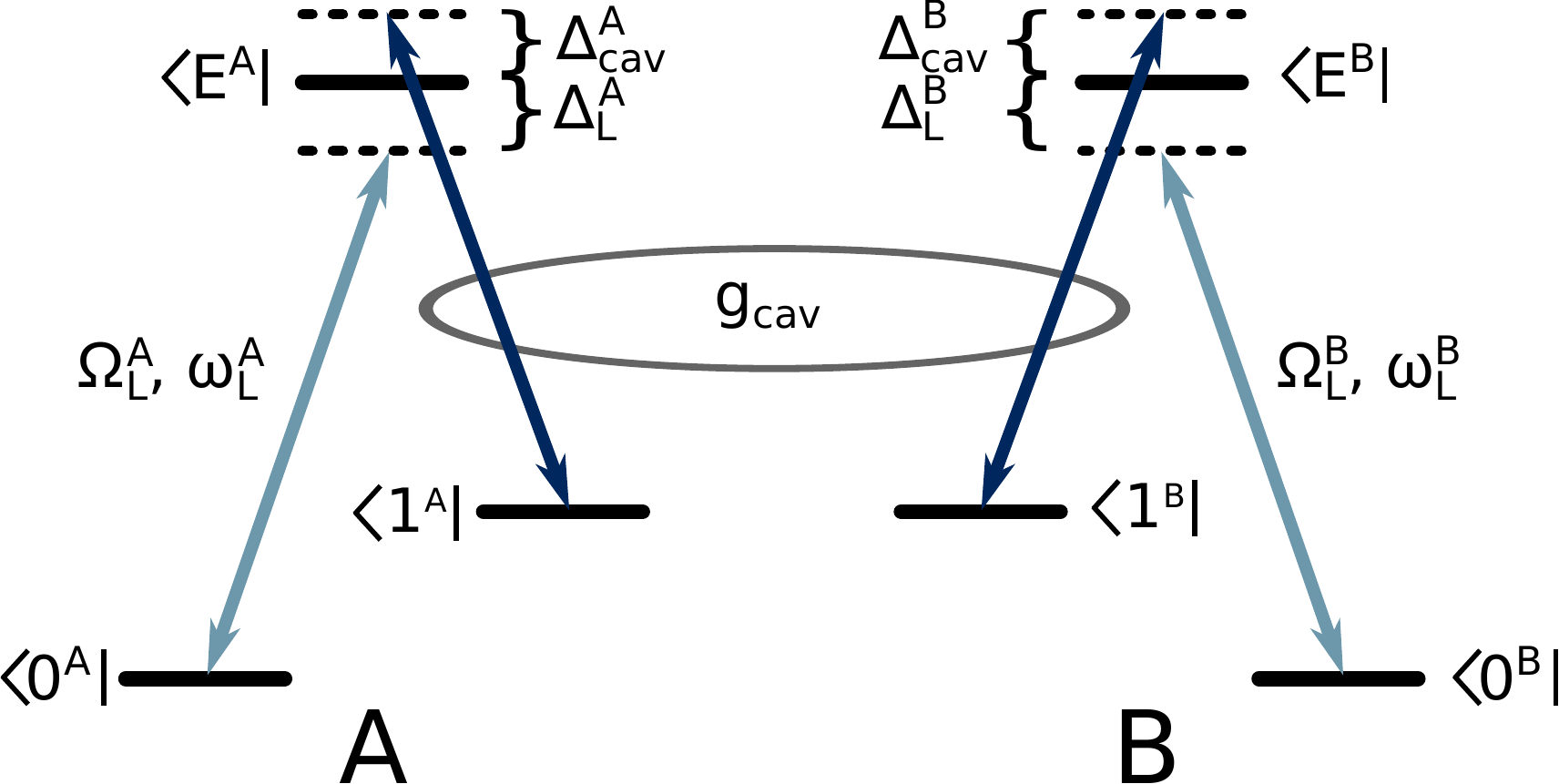}
\caption{(Color online) Level scheme of the two NV centers A and B. 
Each center provides a $\Lambda$-scheme with ground states $\ket{0^{A (B)}}$, $\ket{1^{A (B)}}$ and excited state  $\ket{E^{A (B)}}$. 
The $\ket{0^{A (B)}} \rightarrow \ket{E^{A (B)}}$ transitions are driven by lasers with frequency $\omega_{L}^{A (B)}$ and coupling strength $\Omega_{L}^{A (B)}$ detuned by $\Delta_{L}^{A (B)}$. 
The  $\ket{1^{A (B)}}\rightarrow \ket{E^{A (B)}}$ transition of each system is coupled to the shared cavity mode with coupling strength $g_{cav}^{A (B)}$, where the cavity is detuned by $\Delta_{cav}^{A (B)}$.}
\label{fig:levels}
\end{figure}

A universal two-qubit operation is the spin exchange~\cite{Imamoglu1999}.
For this, one of the Raman lasers is applied to each system, while the second laser is replaced by the cavity mode, as depicted in Fig.~\ref{fig:levels}. 
Importantly, the cavity and the laser detunings $\Delta_{cav}^{A\,(B)}, \Delta_{L}^{A\,(B)}$ are chosen not to match the Raman resonance used in the conventional Raman scheme, i.e. $\Delta_{cav}^{A\,(B)} -\Delta_{L}^{A\,(B)}\neq 0$.  
Now, both system are simultaneously driven by the laser fields and a coherent spin-exchange by stimulated Raman scattering takes place: 
For example system A, initially prepared in $\ket 0$, emits a Raman-photon into the cavity mode, while undergoing a spin flip. 
This process is virtual and can only occur within the time-energy uncertainty, as the photon frequency does not match the cavity resonance.
Only if the photon is absorbed in a second Raman process, where system B undergoes a spin flip, the energy is conserved and the joint spin flip process occurs.

To quantify this, the system of $\ket{1^{A\,(B)}}$, $\ket{0^{A\,(B)}}$ and $\ket{E^{A\,(B)}}$ is described by the Hamiltonian $H=H_{0}+H_{I}$.
When for simplicity assuming equal parameters $\Omega_{L},\omega_{L},\Delta_{cav},\Delta_{L} $ for both $\Lambda$-systems, the free $H_{0}$ and interaction part $H_{I}$ read in the rotating wave approximation
 \begin{eqnarray}
\label{eq:Hamiltonian}
H_{0}&=&\sum_{i,j}\hbar \omega_{i} \ket{i^{j}}\bra{i^{j}} + \hbar\omega_{cav} c^{\dagger}c,\\
H_{I}&=&\sum_{j}\hbar\left[\Omega_{L} \ket{0^{j}} \bra{E^{j}} e^{i\omega_{L}t}+g_{cav}c^{\dagger}\ket{1^{j}}\bra{E^{j}} \right]+\mathrm{h.c.},\nonumber
\end{eqnarray}
where $i\in\{0,1,E\}$, $j\in\{A,B\}$, $g_{cav}$ denotes the cavity coupling for the $\ket{1^{A\,(B)}}\leftrightarrow\ket{ E^{A\,(B)}}$ transition, $\hbar\omega_{i}$ the energy of state $\ket{i^{j}}$ and $c$,$c^{\dag}$ are the usual operators for cavity photons.

\
For a vanishing photon population in the cavity and system $A$ and $B$ initially prepared in the states $|0^A\rangle$ and $|1^B\rangle$ an adiabatic elimination of the excited state manifold as well as the cavity mode leads to an effective interaction between the two spins:
\begin{eqnarray}
\label{Eq:H eff}
H_{\text{eff}}&=&-\hbar\tilde{g}\,\ket{0^{A}}\bra{1^{A}}\otimes \ket{1^{B}}\bra{0^{B}} +\mathrm{h.c.},
\end{eqnarray}
where the effective coupling element is given by:
\begin{eqnarray}
 \tilde{g} = \frac{|\Omega_L|^2 |g_{cav}|^2}{\Delta_L^2(\Delta_{cav}-\Delta_{L}-\frac{2|g_{cav}|^2}{\Delta_L})}.
\label{eq:effective-coupling}
\end{eqnarray}
Later on, by comparing to numerical simulations, we show that our result is much more accurate, compared to previous results from applying 2nd-order perturbation theory after a unitary transformation~\cite{Imamoglu1999}.

\
The time evolution described by the Hamiltonian Eq.~(\ref{Eq:H eff}) is an effective rotation $U_{exc}(\varphi)$ on the 2-spin state  $\ket{S_A,S_B}$.
To generate an entangled state, system A is prepared in the state $\ket{0_{A}}$ while system B is prepared in the state
$\ket{1_{B}}$, i.e. the system of two spins is prepared in the state $\ket{01}$.
Now, by applying a $U_{exc}(\pi/2)$ spin-exchange, this state is transformed into $\ket{\Psi}=1/\sqrt2 (\ket{01} +i\ket{10}$, a maximally entangled state.
This entanglement operation (EO) has three important properties~\cite{Imamoglu1999}: 
1.~It is not necessary, that the two systems are identical. Differences in the optical transition frequency can be compensated by a proper choice of laser frequencies.
2.~Emitters that are detuned from the resonance, i.e \ $\Delta_{L}^{A}+\Delta_{cav}^{A}-\Delta_{L}^{B}-\Delta_{cav}^{B}\neq 0$ \ or outside the laser focus are unaffected, making the mechanism scalable to several systems inside one single cavity. 
3.~By applying single qubit unitary transformations and several spin-exchanges the fundamental c-Not gate can be constructed.

\
In order to realize the EO for a system that is subject to photon decay (with rate $\kappa=\omega_{cav}/Q$) and radiative decay of the excited state manifold $|E^{A(B)}\rangle$ (with rate $\gamma_{rad}$), the following conditions have to be met simultaneously:
\begin{eqnarray}
\label{eq:condition1}
 \Delta_L &\gg&\Omega_L,\\
\label{eq:condition2}
 |\Delta_{cav}-\Delta_L|&\gg& g_{cav},\Omega_L,\kappa,\\
\label{eq:condition3}
\Delta_L^2|\Delta_{cav}-\Delta_L|&\ll&\frac{g_{cav}^2\Omega_L^2}{\gamma_{rad}}.
\end{eqnarray}
Equations \eqref{eq:condition1} and \eqref{eq:condition2} state the limiting conditions for the applicability of $H_{\text{eff}}$. Equation \eqref{eq:condition3} assures a fast spin transfer in comparison to the radiative dephasing $\gamma_{rad}$. While the first two conditions can always be met by sufficiently large $\Delta_L,\Delta_{cav}$, the last constrain sets a potentially contradictory upper bound to the detunings that is determined by $\gamma_{rad}$.
Hence the entanglement scheme cannot be applied to a situation with $g_{cav}\approx\kappa\approx\gamma_{rad}$, as has been realized for atomic~\cite{Weber2009} or quantum dot systems~\cite{Yoshie2004}. In contrast with NV-centers, $\gamma_{rad}$ is significantly smaller compared to $\kappa,g_{cav}$ and Eqs. \eqref{eq:condition1}-\eqref{eq:condition2} can be fulfilled, while Eq. \eqref{eq:condition3} is violated  only weakly. 
In this situation our scheme allows for significant entanglement, even with today's technology. 

\begin{figure}[hbt]
\center
\includegraphics[width=0.49\textwidth]{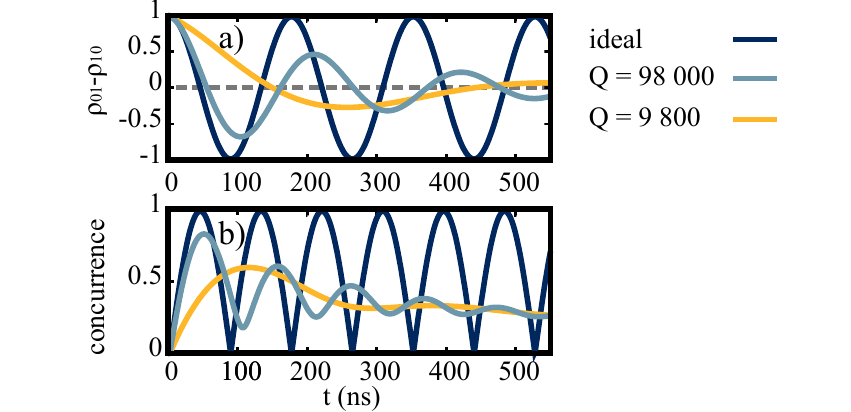}
\caption{(Color online) Dynamics of the two NV spins. 
(a) Calculated inversion $\rho_{01}-\rho_{10}$ between initial state and target state for the ideal case ($\kappa=\gamma=0$) and different Q-factors. 
Even with very moderate Q-factors  significant population transfer is possible. 
(b) Calculated concurrence for (a), indicating the generation of an entangled state during the transfer. Even for a Q-factor as low as $9800$, a high concurrence can be achieved.
For all calculations we used $\Delta_{cav}=9g_{cav}+2\kappa$, $\Delta_{L}=9g_{cav}$, $\Omega_{L}=g_{cav}=2\pi\cdot 3$~GHz.}
\label{fig:3}
\end{figure}
In the following, we verify the analytically predicted applicability of the EO to experimentally feasible 
implementations of NV-center cavity systems. In particular we consider losses from the cavity, radiative dephasing of the excited state $\ket E$ and fluctuations like spectral diffusion.

Including dissipative processes, the equation of motion for the density matrix $\rho$ is given by $d \rho / dt=-i/ \hbar [\rho,H]_{-}+\mathcal L(\rho)$, with the Lindblad form

\begin{eqnarray}
\mathcal L&=&\sum_{x}\hat\gamma_{x}\rho\hat\gamma_{x}^{\dag} -\tfrac{1}{2} [\hat\gamma_{x}^{\dag}\hat\gamma_{x},\rho]_{+}+ \hat\kappa\rho\hat\kappa^{\dag}- \tfrac{1}{2} [\hat\kappa^{\dag}\hat\kappa,\rho]_{+}.
\end{eqnarray}
Here, $x\in\{0^{A},0^{B}, 1^{A}, 1^{B}\}$, $\hat\gamma_{x}=\sqrt{\gamma}\ket{x}\bra{E}$, with $\gamma=50$~MHz describes the decay from the exited stated to ground state $x$ under emission into non-cavity modes and $\hat\kappa=\sqrt{\kappa}c$ losses from the cavity.\\
The equations of motion for the components of the density matrix are expanded and solved using an explicit Runge-Kutta algorithm.
\begin{figure}[b]
\center
\includegraphics[width=0.48\textwidth]{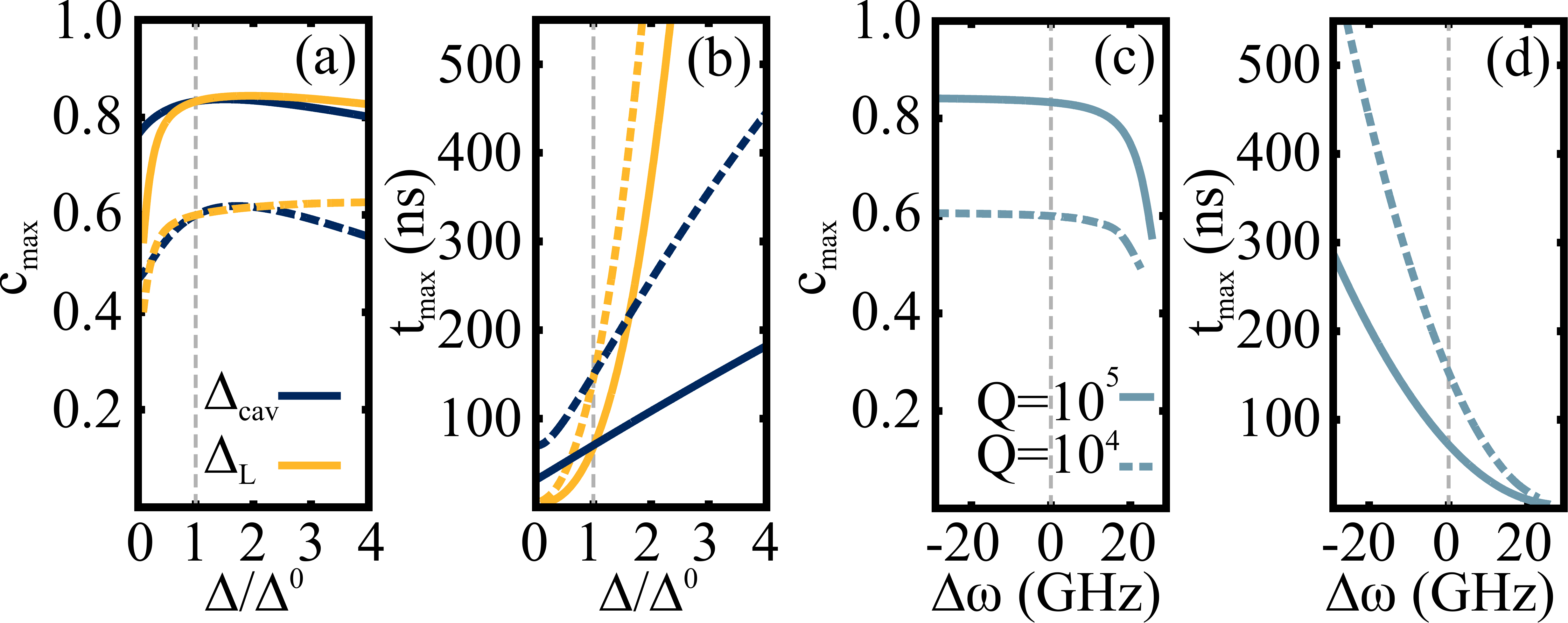}
\caption{(Color online) Influence of the laser and cavity detuning $\Delta_L$, $\Delta_{cav}$ on the entanglement generation.
(a) The achieved maximum concurrence for different Q-factors when varying $\Delta_{L}$ or $\Delta_{cav}$, while keeping the respective other detuning fixed at $\Delta^{0}$.
For the calculations we used $g_{cav}=\Omega=2\pi \cdot 3$~GHz, $\Delta_{L}^{0}=9g_{cav}$, $\Delta_{cav}^{0}=9g_{cav}+2\kappa$, and $Q=10^{4}$ (dashed lines), or $Q=10^{5}$ (solid lines), respectively.
(b) The transfer time needed to achieve the concurrence in (a). (c),(d) Maximum achievable concurrence (c) and EO time (d), when the optical transition frequency changes by $\Delta\omega$.}
\label{fig:4}
\end{figure}
We chose $g_{cav}/(2\pi)=  3.0\,\mathrm{GHz}$ for NV centers localized in the field maximum of a nano cavity~\cite{Mccutcheon2008}. 
This is feasible by slightly improving experimental results on the Purcell enhancement of the zero-phonon transition of NV centers in photonic crystal L3 cavities~\cite{Wolters2010,Faraona}:
With $F=12\,(60)$ being the demonstrated Purcell \mbox{factor~\cite{Wolters2010} (\cite{Faraona})}, $Q=600\,(3000)$ the quality factor of the used cavity, $\tau=14$~ns the lifetime of the excited state, $d=0.05$ the Debye-Waller factor, and $\omega/ (2\pi)=471$~THz the frequency of the optical NV transition we calculate the experimentally achieved coupling to 
\begin{eqnarray}
\frac{g_{cav}}{2\pi}&=&\frac{1}{2\pi}\sqrt{\frac{d \omega F}{4 Q\tau}}\label{eq:purcell2g}=1.15\,\mathrm{GHz}.
\end{eqnarray}
We set $\Omega_{L}=g_{cav}$ which can be achieved even for spin non-preserving transitions with laser powers of about 1~mW~\cite{Santori2006,Robledo2010}. 
To fulfill Eqs. \eqref{eq:condition1}-\eqref{eq:condition3} as well as possible the laser detuning is set to $\Delta_{L}^{0}=9g_{cav}$, while we chose $\Delta_{cav}^{0}=9g_{cav}+2\kappa$ for the cavity detuning.
These values represent a good compromise between radiative dephasing, cavity losses, and time needed for the EO. 
Furthermore, without loss of generality, the ground state splitting is set to the zero field splitting of $\omega_{12}=2\pi\cdot 2.87$~GHz.
With these parameters, we calculate the dynamics for $Q=\omega/\kappa = 9800$, which is in the range of current experiments. 

Starting with NV$^A$ in state $\ket 0^{A}$ and NV$^{B}$ in state $\ket 1^{B}$, i.e. with the diagonal elements $\rho_{ij}\equiv \langle |i,j\rangle\langle i,j|\rangle$ of the density matrix $\rho_{01}=1,\rho_{00}=\rho_{10}=\rho_{11}=0$ a spin exchange takes place, as predicted by the analytical theory.
The maximally achieved inversion is $-(\rho_{01}-\rho_{10})>0.3$, where the  transfer time of  300~ns (Fig.~3) is in agreement with Eq.~(3).
To confirm that the transfer is indeed coherent and an entangled state is prepared, we evaluated the concurrence $c$ \cite{James2001} as a positive definite measure of entanglement during the transfer. A vanishing concurrence indicates a classical, i.e. separable state, while a concurrence of one indicates a maximally entangled state.
Even with the low $Q=9800$, we find a value of $c_{max}\approx0.6$ for the maximally achieved concurrence after the time $t_{max}\approx 150$~ns.
This strikingly demonstrates that even low-Q photonic crystal cavities can mediate entanglement between two NV centers.
When using the challenging, but nevertheless realistic value of $Q=98000$ the EO  even improves.
In this case, we find a maximal inversion of $\rho_{01}-\rho_{10}>0.6$ and a maximal concurrence of $c_{max}\approx0.8$.
\begin{figure}[hbt]
\center
\includegraphics[width=0.49\textwidth]{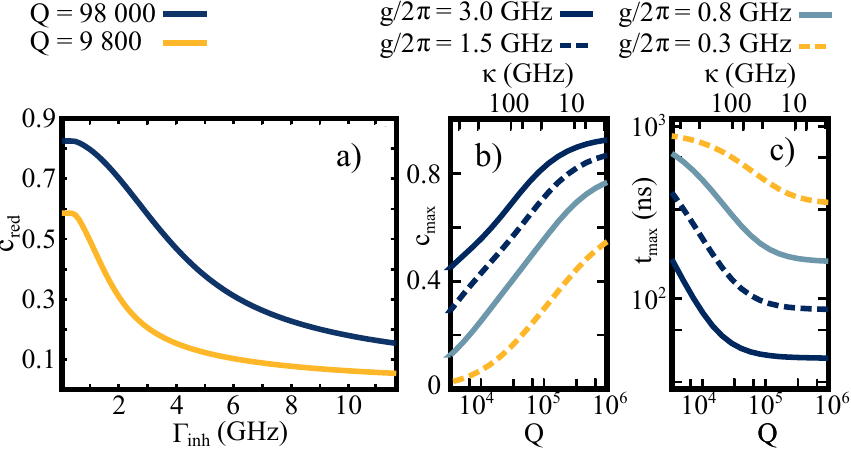}
\caption{(Color online) (a) Expected average concurrence, when the optical transition is broadened to $\Gamma_{inh}$ by spectral diffusion.
(b) The achieved concurrence for different cavity couplings $g_{cav}$ between  $2\pi \cdot 3.0\,\mathrm{GHz}$ and $2\pi\cdot0.3\,\mathrm{GHz}$ when varying the Q-factor.
Even with moderate Q-factors of $10^{4}$ an entangled state can be prepared.
(c) The transfer time required to achieve the concurrence in (c). 
For all calculations we used $\Delta_{cav}=9g_{cav}+2\kappa$, $\Delta_{L}=9g_{cav}$, $\Omega=g_{cav}$.}
\label{fig:5}
\end{figure}

To study the influence of small fluctuations in the laser and cavity detuning,
we calculated the dynamics for varying $\Delta_{L}$ and $\Delta_{cav}$. 
These calculation show that the initial choice of $\Delta_{cav}^{0}$ and $\Delta_{L}$ is indeed a good compromise between efficiency of the EO (Fig. 4(a)) and entanglement time $t_{max}$  (Fig.~4(b)).
The numerical solution shows a linear increase of $t_{max}$  with the cavity detuning $\Delta_{cav}$, while $t_{max}$ depends \textit{quadratically} to \textit{qubic} on the laser detuning at $\Delta_{L}\gg\Delta_{cav}$.
While being in perfect agreement with the effective coupling constant of Eq. \eqref{eq:effective-coupling}, these findings are in clear disagreement with previous analytical results for $ \tilde{g} \sim 1/t_{max}$, where both detunings are predicted to contribute equally and linearly~\cite{Imamoglu1999}.
Hence, our analytic approach is much better suited here.
\

The most important problem in all solid-state systems are fluctuations of the emitters' properties caused by the environment.
For NV centers the optical transition lines jump randomly within a Gaussian envelope of width $\Gamma_{inh}$~\cite{Wolters2013}. These jumps are equivalent to a simultaneous change of the laser detuning by $\pm\Delta\omega$ and the cavity detuning by $\mp\Delta\omega$. Here, the opposite signs guarantee robustness of the EO against spectral diffusion.
Indeed, the achievable concurrence is almost invariant for $\Delta\omega<10\,$GHz (Fig. \ref{fig:4}(c)). Nevertheless, the
EO times changes slightly (Fig. \ref{fig:4}(d)) and dephasing occurs. To study this in more details, we
performed
simulations of the density matrix $\rho(\Delta\omega^{A},\Delta\omega^{B})$ at time $t_{max}$ as a function of the frequency shift  $\Delta\omega^{A (B)}$  with respect to the mean value (Fig.~5(a)).
In an experiment, an average density matrix $\rho(\Delta\omega^{A},\Delta\omega^{B})$ and hence a reduced concurrence $c_{red}$ would be observed, where the actual value of $c_{red}$ depends on the inhomogeneous linewidth $\Gamma_{inh}$.
As a key result of this paper we find that for the realistic case of $\Gamma_{inh}/(2\pi)\sim1$~GHz~\cite{Zhao2013}  the achievable concurrence reaches almost the maximum concurrence, proving the robustness of the EO against spectral diffusion.
\

Finally, in order to investigate the influence of the cavity quality factor $Q$ and coupling $g_{cav}$ in detail, we calculated the maximum achievable concurrence $c_{max}$ (Fig. 5(b)) and needed entanglement time $t_{max}$ (Fig. 5(c)) as a function of Q for various couplings between $g_{cav}=2\pi\cdot 0.3$~GHz and $g_{cav}=2\pi\cdot 3.0$~GHz. 
As expected, for small Q-factors photon loss from the cavity modes limits the achievable concurrence. 
Furthermore, a strong dependency on the coupling constant $g_{cav}$ is visible.
This can be explained by Raman scattering into non-cavity modes that induces additional unintended spin flips and dominates the dynamics for low ratios between $g_{cav}$ and $\gamma$.

In conclusion, small mode volume photonic crystal cavities with comparably low Q-factors can be an important tool on the path towards deterministic entanglement of medium distant NV centers.
This opens the way for future quantum information processing networks under realistic conditions, i.e. including unavoidable fluctuations, such as spectral diffusion.
Future work will be devoted to the improvement of the entanglement scheme via pulse shaping and detailed parameter analysis.
Furthermore, the prospects of adjacent nuclear spins for the use as local quantum registers will be investigated.

This work was supported by the DFG (FOR 1493, SFB 787 and SFB 910 (J.K.)).
J.W. acknowledges funding by Humbold-Universi\"{a}t zu Berlin (Humbold Post-Doc Scholarship). 

\bibliographystyle{unsrt}

\end{document}